# Weakly Supervised Lesion Detection and Diagnosis for Breast Cancers with Partially Annotated Ultrasound Images

Jian Wang, Liang Qiao, Shichong Zhou, Jin Zhou, Jun Wang, *Member*, *IEEE*, Juncheng Li, Shihui Ying, *Member*, *IEEE*, Cai Chang, and Jun Shi, *Member*, *IEEE*

*Abstract*—Deep learning (DL) has proven highly effective for ultrasound-based computer-aided diagnosis (CAD) of breast cancers. In an automatic CAD system, lesion detection is critical for the following diagnosis. However, existing DL-based methods generally require voluminous manually-annotated region of interest (ROI) labels and class labels to train both the lesion detection and diagnosis models. In clinical practice, the ROI labels, i.e. ground truths, may not always be optimal for the classification task due to individual experience of sonologists, resulting in the issue of coarse annotation that limits the diagnosis performance of a CAD model. To address this issue, a novel Two-Stage Detection and Diagnosis Network (TSDDNet) is proposed based on weakly supervised learning to enhance diagnostic accuracy of the ultrasound-based CAD for breast cancers. In particular, all the ROI-level labels are considered as coarse labels in the first training stage, and then a candidate selection mechanism is designed to identify optimal lesion areas for both the fully and partially annotated samples. It refines the current ROI-level labels in the fully annotated images and the detected ROIs in the partially annotated samples with a weakly supervised manner under the guidance of class labels. In the second training stage, a self-distillation strategy further is further proposed to integrate the detection network and classification network into a unified framework as the final CAD model for joint optimization, which then further improves the diagnosis performance. The proposed TSDDNet is evaluated on a B-mode ultrasound dataset, and the experimental results show that it achieves the best performance on both lesion detection and diagnosis tasks, suggesting promising application potential.

## I. INTRODUCTION

BREAST cancer is one of the most common cancers that seriously threatens women's health. B-mode ultrasound (BUS) is a routine imaging tool for diagnosing breast cancers in clinical practice [1]. With the fast development of deep learning (DL) techniques, the BUS-based computer-aid diagnosis (CAD) has gained its reputation in recent years, and it can help sonologists to improve diagnostic accuracy together with consistency and repeatability [1].

An automatic BUS-based CAD system mainly comprises two fundamental modules, namely breast lesion detection and classification [1]. The former automatically detects the potential lesions and localizes the corresponding regions of interest (ROIs), while the latter differentiates between benign tumors and malignant cancers conducted on the detected ROIs [2]. Therefore, lesion detection is a critical module that directly influences subsequent diagnostic accuracy in the classification module. Existing DL-based detection methods generally require large numbers of labeled samples for training the detection network [3]. However, ROI annotation is time-consuming and laborious for sonologists, which is one of the key causes for the small sample size (SSS) problem. Therefore, limited training samples will significantly affect the performance of breast lesion detection models.

On the other hand, the manually annotated ROIs by sonologists are often used as ground truth in training the detection model. However, they may not always be optimal for subsequent classification task, since it is often difficult to define a uniform criterion for ROI-level annotations. In fact, this annotation heavily relies on the sonologist's experience [6]. As shown in Fig. 1, the ROI bounding boxes with different sizes have different prediction probabilities on the same breast lesion, significantly affecting the classification performance [6][7]. Specifically, a large-size ROI bounding box contains too much redundant information, while the small-size one only includes limited lesion region and may miss some crucial diagnostic information, such as posterior acoustic features [8]. Both cases can then degrade the diagnostic accuracy of a CAD model. Therefore, a medium-size bounding box is encouraged to strike a balance between large-size and small-size bounding boxes with superior performance.

In this work, we refer to this uncertainty in ROI annotation as the issue of *coarse annotation* according to references [9] and [10]. Additionally, Yamakawa et al. [11] also noted this issue for the ultrasound-based diagnosis of liver tumors. They adopted the ratio between the maximum diameter of the tumor and the ROI size as the index for ROI cropping, and concluded

This work is supported in part by the National Key Research and Development Program of China (2021YFA1003004), National Natural Science Foundation of China (81830058, 62271298, 11971296) and the 111 Project (D20031). (Corresponding authors: Jun Shi)

J. Wang, L. Qiao, J. Wang, J. Li and J. Shi are with the Key Laboratory of Specialty Fiber Optics and Optical Access Networks, Joint International Research Laboratory of Specialty Fiber Optics and Advanced Communication, Shanghai Institute for Advanced Communication and Data Science, School of Communication and Information Engineering, Shanghai University, China. (Email: junshi@shu.edu.cn)

S. Ying is with the Department of Mathematics, School of Science, Shanghai University, China

S. Zhou, J Zhou, and C. Chang are with the Fudan University Shanghai Cancer Center, Fudan University, China.

that the model achieved optimal performance when the index was set to 0.6 through experiments. Consequently, the coarse annotation is another factor affecting lesion detection and diagnosis performance.

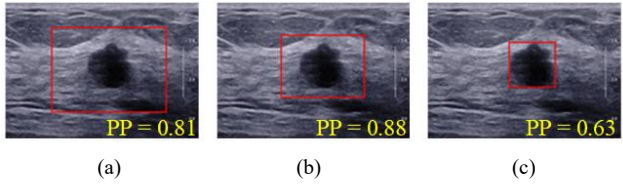

Fig. 1 Three different ROI size cases with their prediction probability on a same BUS image by ResNet18. (a) large-size, (b) medium-size, and (c) small-size. PP means prediction probability.

Semi-supervised learning (SSL) is a natural way to train a lesion detection model by utilizing both annotated and unannotated ROIs in BUS images. This method is certain to achieve superior detection performance compared to the supervised learning with only a small number of labeled samples [10]. However, SSL cannot guarantee the accuracy of the ROI-level pseudo labels predicted during model training, because there is generally a lack of effective performance criteria for evaluating and correcting detected lesion regions during the training procedure [12][13][14].

It is worth noting that all the training samples have image-level disease labels, i.e., benign tumor and malignant cancer, for training a CAD model, since they are the retrospective data in clinical practice. However, they may be not completely annotated with ROIs by sonologists because of the time-consuming and laborious annotation. Thus, we split the training images into two groups: one containing fully labeled images with both ROI-level and image-level labels, and another including partially annotated images with only image-level labels but without ROI-level labels. We argue that these image-level class labels can potentially enhance lesion detection during model training [14][15].

Weakly supervised learning (WSL) is a feasible method for applying these image-level labels to locate more accurate ROIs for the following classification task. We argue that the classification accuracy can be used as a criterion to adjust the location and size of automatically detected ROIs. In addition, this idea can also solve the issue of coarse annotation. That is, the manually annotated ROIs can be refined according to the corresponding classification accuracies by slightly changing their localizations and sizes. To the best of our knowledge, there is no relevant research on the WSL-based ROI refinement in the field of medical image analysis, especially not for refining manual ROI-level labels (ground truth).

In this work, a WSL-based Two-Stage Detection and Diagnosis Network (TSDDNet) is proposed for the BUS-based CAD of breast cancers, which can effectively solve the issues of coarse annotation and improve both the performance of lesion detection and classification with limited training samples. It consists of a lesion detection Network (D-Net) and a classification network (C-Net). In the first training stage, the D-Net and C-Net are jointly trained through a WSL-based ROI refinement procedure. In the second training stage, a joint optimization with the self-distillation strategy is developed to integrate the D-Net and C-Net into a unified framework, which then further promotes the detection and classification performance of the overall CAD model. The experimental results on a B-mode ultrasound dataset indicate the effectiveness of the proposed TSDDNet.

The main contributions of this work are three-fold:
1) A WSL-based ROI refinement method is developed to not only improve detection accuracy of the ROI-level pseudo labels predicted by the detection network, but also refine the manually annotated ground truth ROIs. Specifically, to the best of our knowledge, it is the first work to refine the existing ROI-level labels (ground truth) during the stage of model training by a specially designed candidate selection mechanism.
2) A novel TSDDNet is proposed for automatically detecting breast lesion and then diagnosing in a unified framework. Furthermore, a two-stage training strategy is designed so that the TSDDNet can be jointly trained and optimized with both the fully and partially annotated images for improving its performance.
3) The self-distillation based joint optimization is proposed to incorporate the D-Net and C-Net into a unified framework in the second training stage. The self-distillation strategy is designed to transfer the knowledge from the latter classification task to the former lesion detection task with the improved performance of the overall CAD model.

## II. RELATED WORK

### A. BUS-based CAD for Breast Cancers

Breast lesion detection is a critical step in an automatic BUS-based CAD system [16]. Some DL-based approaches have been developed for this task. For example, Yap et al. compared three DL-based models, i.e., LeNet, U-Net, and FCN-AlexNet, for lesion detection [18], among which FCN-AlexNet achieved the best performance; Zhang et al. proposed a breast lesion detection network by introducing a Bayesian model into YOLOv4 [19]. Meanwhile, DL has also been the mainstream method in the BUS-based CAD models for breast cancers. For example, Qi et al. designed a convolutional neural network (CNN) with multi-scale kernels to improve diagnostic accuracy [2]; Moon et al. utilized several CNN architectures to generate different image content representations and fused them to develop a CAD system for tumor diagnosis.[20]. These works demonstrate that DL can achieve promising detection and diagnosis performance for the BUS-based CAD of breast cancers, but the detection and classification tasks are generally implemented in individual systems.

To this end, some CAD models integrate lesion detection and classification tasks into a unified framework. For example, Huang et al. employed ROI-CNN and G-CNN to construct a two-stage grading system for automatic diagnosis of breast tumors, in which the ROI-CNN was used for breast lesion detection, and the G-CNN was utilized for subsequent classification [22]; Shin et al. constructed a framework to simultaneously localize and diagnose tumors with BUS images, which could automatically select loss function for weakly and semi-supervised training scenarios, respectively [23]. These works indicate that the unified CAD framework often achieves superior performance over the individual models.

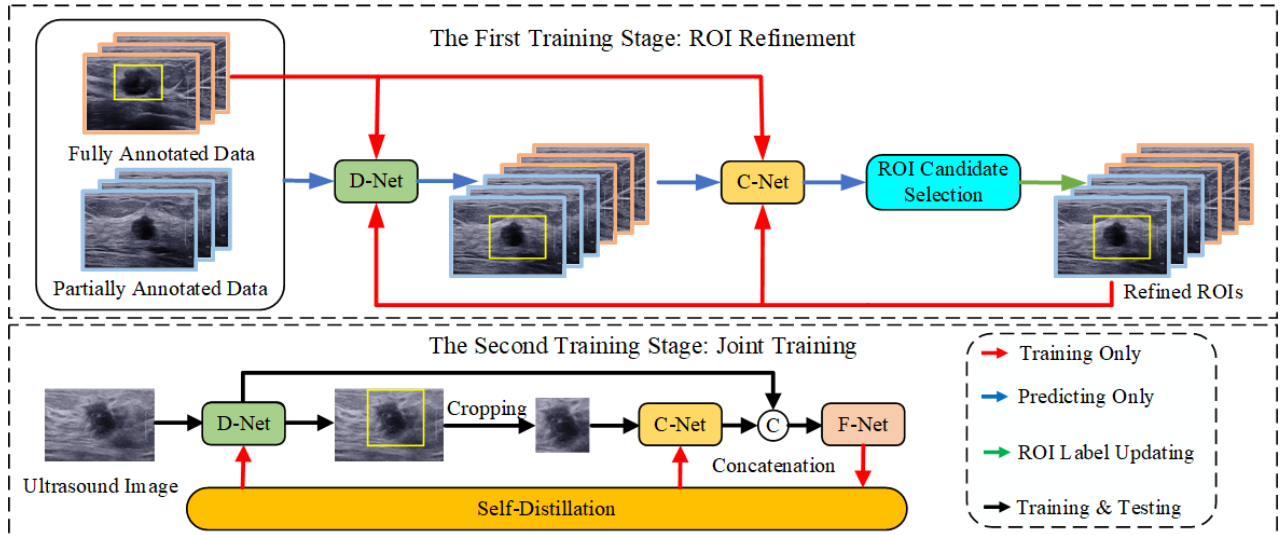

Fig. 2 Framework of the proposed TSDDNet. In the first stage, fully annotated data is used to train both the C-Net and D-Net, and then partially annotated data is fed to the trained D-Net to generate pseudo-ROIs. A candidate selection strategy is designed to refine ROI-level labels of both fully and partially annotated data during multiple iterations. In the second stage, self-distillation strategy is used to further finetune the D-Net and C-Net to promote classification performance.

In this work, we also develop a unified approach, named TSDDNet, to not only alleviate the problem of SSS by the WSL with unannotated images, but also address the issue of coarse annotation by design a candidate selection mechanism to refine both the pseudo and manually annotated ROI-level labels. Meanwhile, a self-distillation strategy is developed to further promote the classification performance of the TSDDNet.

### B. WSL for Lesion Detection

Due to the high cost and tedious process of medical image annotation, it is generally difficult to collect a large number of annotated samples to train the object detection model. Recently, the WSL-based object detection method has attracted considerable attention [27], including for detecting lesions in medical images. It can train a detection network with only image-level labels [28][32]. For example, Hwang et al. developed a two-stream CNN model to localize the tuberculosis regions in chest X-ray images by leveraging class activity maps to generate pseudo-ROIs [29]; Dubost et al. proposed an encoder-decoder architecture to compute high resolution attention maps through segmentation features generated by image-level labels [30]; Shin et al. adopted a weakly annotated dataset with image-level labels and a smaller strongly annotated dataset with both image-level labels and ROIs in a mixed manner to develop a joint WSL and SSL-based CAD for breast cancers [23]; Kim et al. utilized class activity maps generated by three classification networks to detect and diagnose breast cancer without image annotation [31]. All these works indicate the effectiveness of the weakly supervised lesion detection.

However, existing works always take the manually annotated ROIs as ground truth to train the lesion detection model, which results in the problem of coarse annotation and affects subsequent classification tasks. Therefore, we propose a novel WSL-based CAD model with a candidate selection mechanism to refine the manually annotated ROIs to be more suitable for the classification task.

### C. Self-distillation

Self-distillation aims to distill knowledge within a network itself, which firstly divides the network into several sections and then squeezes the knowledge of the deeper portion into the shallow ones [35]. Recently, it has been introduced to promote model performance in various computer vision tasks[33]. For example, Zhang et al. proposed the self-distillation algorithm to avoid consuming too much time on training teacher model and searching student model[35]; Hou et al. presented a distillation model to enhance the CNN-based lightweight lane detection, which utilized attention maps on the deeper layers to distillate lower layers [36]; Yang et al. developed a snapshot distillation, which transferred knowledge from the earlier epochs of the training process of the network into later epochs to boost the performance of CNN [37]. Luo et al. developed a self-distillation augmented masked autoencoders to enhance the feature representation on top of autoencoders for histopathological image classification [38]. In this work, a self-distillation strategy is introduced into TSDDNet to promote both the lesion detection and classification networks.

## III. METHODOLOGY

### A. Two-Stage Detection and Diagnosis Network

As shown in Fig. 2, the proposed TSDDNet consists of three sub-networks: a detection network D-Net, a classification network C-Net, and a fusion network F-Net.

A WSL-based two-stage training strategy is developed for this new detection and diagnosis network. Specifically, in the first training stage, both D-Net and C-Net are trained using the fully annotated images with both ROI-level and image-level labels, respectively. After that, the partially annotated images with only image-level labels are fed into the trained D-Net to generate ROI-level pseudo labels. Moreover, a candidate selection mechanism is designed to refine the ground truth ROI labels in the fully annotated images and the pseudo-ROI labels in the partially annotated images. In the second training stage, a self-distillation method is adopted to further finetune the D-Net and C-Net by squeezing knowledge from F-net into the D-Net and C-Net.

It is worth noting that C-Net plays different roles in the two training stages. In particular, in the first stage, it generates the classification probability for each ROI candidate. While in the second stage, C-Net is adopted to extract discriminative feature representation of ROI images. After learning features from both D-Net and C-Net, F-Net is used to predict the final result by fusing these features. Meanwhile, the two classifiers used for self-distillation in the last layer of D-Net and C-Net will be removed in the testing phase.

In order to build an elegant model, the architectures of D-Net, C-Net, and F-Net are specially designed according to their roles, respectively.

**D-Net.** The D-Net is designed based on RetinaNet for lesion detection from BUS images [39], but it only retains the localization branch and removes the classification branch. As shown in Fig. 3, the D-Net is composed of a backbone network and three detection-specific head networks. The Feature Pyramid Network (FPN) is selected as the backbone for integrating multi-scale feature maps [40], and the three head networks perform bounding box regression in the form of multi-scale convolution. Meanwhile, the pseudo-ROIs generated by D-Net are served as ROI candidates for the following candidate selection.

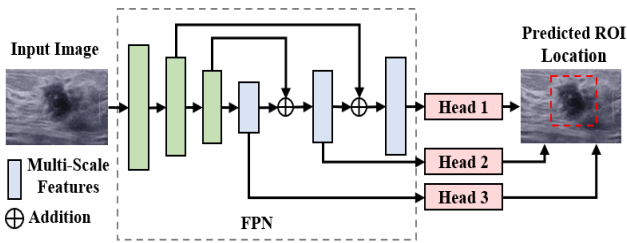

Fig. 3 Architecture of D-Net with Feature Pyramid Network (FPN) as backbone.

**C-Net.** The C-Net is an individual classification network with a classifier that adopts the structure of ResNet as backbone [41]. In the first training stage, the C-Net modifies the final fully connection layer of ResNet for the binary classification task. While in the second training stage, the fully connection layer of ResNet is replaced with a classifier for self-distillation. In the first and second training stages, C-Net can generate discriminative features for subsequent classification task.

**F-Net.** Although the C-Net can directly give the predictive result, we still design a F-Net to further perform a more robust classification. That is, F-Net integrates the location features and discriminative features from D-Net and C-Net, respectively, into a simple network for the final classification. As shown in Fig. 5, F-Net is composed of two fully connection layers and a softmax function.

### B. The First Training Stage: Weakly Supervised ROI Detection and Refinement

In this work, we divide the training samples into two groups. One includes fully annotated images that have both ROI-level and image-level labels, and the other one includes partially annotated images that only have image-level labels.

We first train both the D-Net and C-Net with the fully annotated images. Thereafter, the D-Net is applied to the fully and partially annotated images to generate the pseudo labels, i.e., ROIs for lesions. Both the ROIs in the fully and partially annotated images are fed into the C-Net to get the classification score for each ROI, which indicates the possibility of the lesion being benign or malignant.

Meanwhile, the classification score serves as the criterion to decide whether to replace the current ROI-level label in a BUS image with the newly generated ROI by D-Net. That is, the ROI with high score will replace the previous one with low score. By iterating the above process for $k$ times, ROIs with high scores can constantly replace previous ROIs with low scores in the fully and partially annotated images. Therefore, the most suitable ROIs for classification tasks are selected from $k$ ROI candidates. Finally, we obtain refined ROI-level labels for the fully and partially annotated images and both the D-Net and C-Net are well trained to locate and classify lesions, respectively.

**Candidate Selection Mechanism.** In the proposed TSDDNet, the original ROI-level labels in the dataset are viewed as coarse annotations, which may affect the subsequent classification performance. Therefore, we suggest to employ class labels as prior weakly supervised information to guide the refinement for not only the pseudo-ROIs but also the ground truth ROIs. Fig. 4 shows the flowchart of the candidate selection mechanism in the first training stage. The bounding boxes predicted by D-Net are used as ROI candidates to search for more suitable lesion regions for classification in BUS images. Meanwhile, a hyperparameter $k$ is introduced to control the number of candidates. The C-Net is then applied to compute the classification probability of each ROI candidate, and the candidate with the highest probability is selected to replace the current ROI-level label. To the best of our knowledge, this is the first work in the field of medical images that utilizes class labels as weakly supervised information to guide ROI detection, and specifically refine the ground truth ROIs.

**Optimization.** In the first training stage, a localization loss is adopted to optimize the D-Net, which contains two parts, i.e., the weighted localization losses on the fully annotated (*fa*) images and partially annotated (*pa*) images, respectively:

$$L_{D-Net} = L_{reg}^{roi}(fa) + \alpha L_{reg}^{roi}(pa) \quad (1)$$

where $L_{reg}^{roi}(fa)$ and $L_{reg}^{roi}(pa)$ denote the localization losses of the fully annotated images and partially annotated images, respectively, and $\alpha$ is a hyper-parameter to control the contribution of localization loss in the *pa* set. Meanwhile, the standard smooth $L_1$ loss is employed to compute the localization loss $L_{reg}^{roi}$ as follows:

$$L_{reg}^{roi} = \sum_{i\in\{x,y,w,h\}} smooth_{L_1}(t_i - v_i) \quad (2)$$

where $t_i$ is the coordinates of predicted ROI location and $v_i$ represents the ground-truth ROI box associated with a positive anchor.

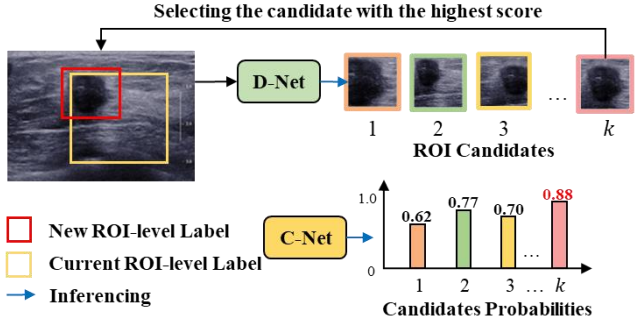

Fig. 4 Candidate Selection mechanism. The BUS images are fed to the D-Net to predict $k$ candidates and the C-Net evaluates the probabilities on each candidate, selecting the candidate with the highest probability as the new ROI-level label.

Similar to the D-Net, the loss function of the C-Net is defined as the weighted classification loss on the *fa* and *pa* sets:

$$L_{C-Net} = L_{cls}^{roi}(fa) + \beta L_{cls}^{roi}(pa) \quad (3)$$

where $L_{cls}^{roi}(fa)$ and $L_{cls}^{roi}(pa)$ denote the classification losses of fully annotated data and partially annotated data, respectively, and $\beta$ is a hyperparameter that controls the contribution of classification loss in the *pa* set. In this part, the $L_{cls}^{roi}$ is the softmax cross-entropy loss. The output of the classification network's softmax layer is represented as a set $Q = \{q_i\}_{i=1}^N$ of predicted class probabilities, where $i$ is the index of a sample and $N$ is the total number of samples. Correspondingly, the set of class labels is denoted as $Y = \{y_i\}_{i=1}^2$, where $y_i \in \{0,1\}$ represents whether a given sample belongs to a benign or malignant lesion. The $L_{cls}^{roi}$ is computed as follows:

$$L_{cls}^{roi} = -\frac{1}{N}\sum_{i=1}^{N}[y_i \log(q_i) + (1-y_i)\log(1-q_i)] \quad (4)$$

where $q_i$ represents the softmax layer's output of the $i$-th sample and $y_i$ is the class label of the $i$-th sample.

### C. The Second Training Stage: Joint Training Based on Self-Distillation

In the first training stage, the class labels have been utilized to guide ROI refinement by WSL. Therefore, the detected ROIs have been refined to be more suitable for the following classification task. However, the D-Net and C-Net are two independent networks, leading to insufficient integration of discriminative features and lesion location features generated by C-Net and D-Net, respectively. To this end, an additional F-Net together with a self-distillation based joint training strategy is developed in the second training stage to further promote the performance of classification. As shown in Fig. 5, the F-Net integrates the features from both D-Net and C-Net to enhance feature representation for the classification task. Meanwhile, the self-distillation strategy is adopted to further improve the performance of C-Net and D-Net.

**Self-Distillation Strategy.** To further enhance the diagnostic accuracy, a self-distillation strategy is employed to strengthen feature extraction capability of both C-Net and D-Net. As illustrated in Fig. 5, the three sub-networks are cascaded in the second stage of TSDDNet. Meanwhile, the fully connection layer in C-Net is removed and two classifiers are added after the last layer of the D-Net and C-Net, respectively, during the joint training stage. Specifically, the classification information in the deep portion (F-Net) is squeezed into the shallow ones (C-Net and D-Net) to improve the lesion localization and discrimination performance of C-Net and D-Net, respectively.

**Optimization.** In order to optimize TSDDNet, the overall loss function of TSDDNet in the second training stage is formulated as the sum of four loss functions:

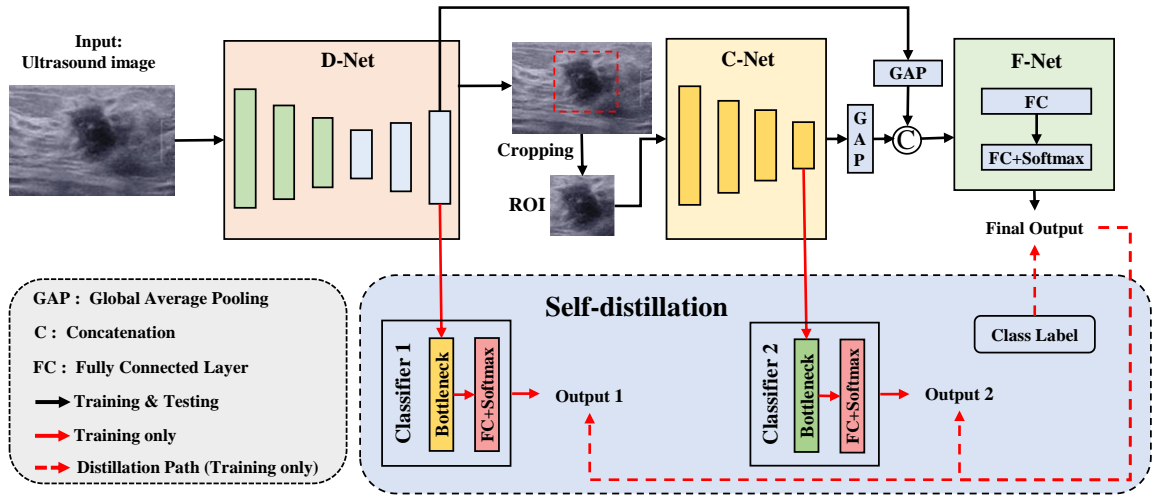

Fig. 5 Joint training with self-distillation strategy. The self-distillation strategy consists of two classifiers and each of them connects the last convolutional layer in the D-Net and C-Net, respectively. During the training, the D-Net and C-Net with corresponding classifiers are trained as student models via distillation from the F-Net to further improve the classification performance.

$$L_{j-train} = L_{D-Net} + L_{F-Net} + L_{cls1} + L_{cls2} \quad (5)$$

where $L_{D\text{-}Net}$ is a localization loss of the D-Net in Eq. (2), $L_{F\text{-}Net}$ is the classification loss for F-Net, and $L_{cls1}$ and $L_{cls2}$ are the classification losses for the two additional classifiers, respectively. The $L_{F\text{-}Net}$, $L_{cls1}$, and $L_{cls2}$ are the same as Eq. (4).

It is worth noting that the localization loss in D-Net is still preserved in the second training stage. Since the ROI-level labels have been refined to fit the classification task in the first training stage, they can indicate the optimal lesion area in the second training stage. Therefore, minimizing the localization loss is also beneficial to improve the classification results.

## IV. EXPERIMENTS AND RESULTS

### A. Datasets

To evaluate the effectiveness of the proposed TSDDNet algorithm, we conducted experiments on a B-mode breast ultrasound image (BBUI) dataset acquired from Fudan University Shanghai Cancer Center. The approval from the ethics committee of the hospital was obtained, and all patients signed informed consent.

All BUS images were scanned from 176 patients (89 patients with benign tumor and 87 patients with malignant cancer), and each patient had 10 BUS images extracted from their ultrasound scanning videos. Thus, there were totally 890 benign images and 870 malignant samples. All samples were scanned by the Mindary Resona7 ultrasound scanner (Shenzhen Mindray Bio-Medical Electronics Co., Ltd., Shenzhen, China) with the L11-3 linear-array probe. For each BUS image, a rectangle ROI was annotated by an experienced sonologist to indicate the lesion area.

### B. Experimental Setup and Evaluation Metrics

The proposed TSDDNet was compared with the following related algorithms:
1) RetinaNet [39]: The RetinaNet was selected as the baseline for the detection and classification tasks in the CAD.
2) Faster R-CNN [42]: Faster R-CNN is a classical supervised object detection framework with ResNet as backbone, which employs a region proposal network (RPN) to generate ROIs.
3) YOLOv4 [17]: YOLOv4 is a real-time detection framework that can predict both the bounding box coordinates and class probabilities. Here, the ResNet was used as the backbone for a fair comparison in this work.
4) YOLOv7 [43]: YOLOv7 is the newest version of the YOLO series algorithms, which was also compared.
5) STAC [44]: STAC was selected for comparison as the classical SSL-based detection and classification, in which the self-training and augmentation-driven consistency regularization were adopted to generate pseudo-ROIs on image-level data.
6) Unbiased Teacher [45]: It is an SSL-based approach that jointly trains a Student model and a gradually progressing Teacher model in a mutually beneficial manner for detection and classification.
7) Soft Teacher [46]: It is an end-to-end SSL-based algorithm, which adopts a box jittering approach to select reliable pseudo-ROIs.
8) SPA [48]: Structure-Preserving Activation (SPA) is a two-stage WSL-based approach, which develops a structure-preserving activation to fully leverage the structure information for localizing objects.

Furthermore, an ablation experiment was conducted to compare TSDDNet with the following variants:
1) TSDDNet-B: This variant adopted the same network structure as TSDDNet, but removed both the candidate selection mechanism and self-distillation strategy. It worked as a basic model of the two-stage framework in the ablation experiment.
2) TSDDNet-B+CS: This variant utilized the same network as TSDDNet-B, but only applied the candidate selection mechanism in the first training stage to refine the ROI-level labels, and did not conduct self-distillation strategy in the second training stage.
3) TSDDNet-B+SD: This variant also adopted the same network as TSDDNet-B, but only performed the self-distillation strategy in the second training stage, and did not conduct the candidate selection mechanism in the first training stage.

The 5-fold cross validation was performed to evaluate all the algorithms. In each fold, the entire BBUI dataset was divided into 70%, 10%, and 20% for training, validation, and testing. It had been ensured that no overlapped patients existed across the three splits. In addition, for the weakly supervised detection, a parameter $p$ was introduced to control the percentage of samples with ROI-level labels in the training dataset. For example, when $p$ was set to 0.2, it indicated that 20% of the patient samples in the training set were randomly selected to retain ROI-level labels, while the rest kept only the class labels.

The commonly used classification accuracy, sensitivity, specificity, and Youden index (YI) were selected as evaluation metrics. Moreover, the receiver operating characteristic (ROC) and the area under the ROC curve (AUC) were also utilized for evaluation.

### C. Implementation Details

The standard ResNet-34 equipped with FPN was used to construct the backbone of the D-Net to extract multi-scale localization features of ROIs. Meanwhile, anchors with 3 scales and 3 aspect ratios of the D-Net were as same as those in RetinaNet [39]. In addition, the Resnet-18 was adopted as the backbone of the C-Net to extract the classification features in ROI images [41].

The ROI images generated by D-Net were resized to 224×224 pixels and then fed into the C-Net. In the first training stage, the Adam optimization algorithm was adopted with the learning rate 1e-4. The minibatch size was 8, and the $k$ was set to 10. Both hyperparameters $\alpha$ and $\beta$ were set as 0.8. Moreover, the backbone parameters of the D-Net and C-Net have been initialized with the corresponding models pretrained on the ImageNet dataset. In the second training stage, the Adam optimization algorithm was also employed with the learning rate 1e-5. The minibatch size was 16, and the *k* was set to 1. Both hyperparameters *α* and *β* were set as 1.0. Furthermore, the weights of the D-Net and C-Net trained in the first training stage were utilized for initialization.

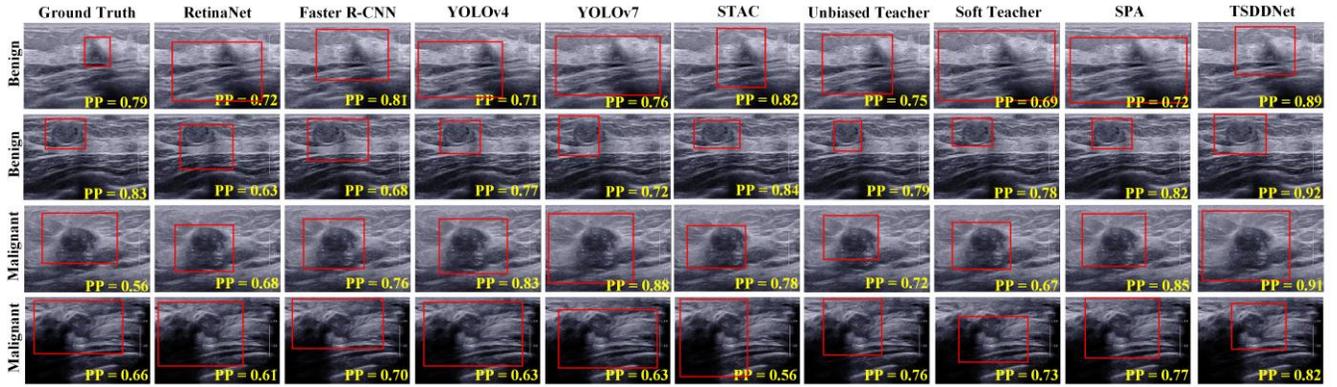

Fig. 6 Visualization results of manual and predicted ROIs. PP means Predicted Probability.

## V. EXPERIMENTAL RESULTS

### A. Visualization Experiment

To verify the effectiveness of the ROI refinement, both the manual ROI-level labels and predicted ROIs are visualized. The predicted lesions were generated based on the coordinates produced by different algorithms, including RetinaNet [39], Faster R-CNN [42], YOLOv4 [17], Yolov7 [43], STAC [44], Unbiased teacher [45], SPA [48], Soft Teacher [46], and the proposed TSDDNet. The predicted probabilities are computed by the C-Net to evaluate the classification performance of the refined ROIs.

As shown in Fig. 6, TSDDNet achieves the best detection results on breast lesions with the highest classification probability in BUS images, indicating that the ROIs predicted by TSDDNet are more suitable for the subsequent classification task in CAD. This benefits from the candidate selection mechanism used in TSDDNet. Specifically, in each iteration of the first training stage, the ROI-level labels of the training samples are refined according to their classification performance. In the second training stage, the classification performance is further enhanced by the self-distillation strategy. Therefore, TSDDNet outperforms the compared algorithms, including RetinaNet, Faster R-CNN, YOLOv4 Yolov7, STAC, Unbiased teacher, SPA, and Soft Teacher. In addition, as a WSL-based approach, TSDDNet also indicates the potentially optimal ROI-level label for BUS images.

### B. Comparison Experiments

Table I shows the results of the breast cancer classification performance of different comparison algorithms. The weakly supervised approaches used a p-value of 0.2, while the supervised algorithms used a p-value of 1.0. The proposed TSDDNet achieves the best average accuracy of 89.62 ± 1.24%, sensitivity of 90.73±1.15%, specificity of 87.59±1.27%, and YI of 78.32±1.95%. It also gets the improvements by at least 3.53%, 3.47%, 2.69%, and 6.59% on the corresponding indices over other compared algorithms. The results suggest that the two-stage training strategy of TSDDNet can improve the classification performance, and the refined ROI can provide the network with more discrimination information to classify lesion types.

TABLE I
CLASSIFICATION RESULTS OF DIFFERENT METHODS ON BBUI DATASET (UNIT: %)

| Method | Accuracy | Sensitivity | Specificity | YI |
|---|---|---|---|---|
| RetinaNet | 84.84±1.08 | 86.01±0.82 | 82.70±1.09 | 68.71±1.74 |
| Faster R-CNN | 85.65±1.27 | 86.04±1.31 | 83.74±1.28 | 69.78±2.13 |
| YOLOv4 | 84.93±1.21 | 85.17±1.46 | 82.46±1.23 | 67.63±2.38 |
| YOLOv7 | 86.09±1.14 | 87.26±1.35 | 84.46±1.46 | 71.72±2.23 |
| STAC | 82.42±1.70 | 83.33±1.13 | 81.34±1.07 | 64.67±1.99 |
| Unbiased teacher | 83.82±1.13 | 84.73±1.13 | 82.74±1.62 | 67.47±1.59 |
| Soft teacher | 85.02±1.35 | 85.85±1.25 | 83.96±1.27 | 69.81±2.02 |
| SPA | 85.77±1.30 | 86.83±2.40 | 84.90±2.65 | 71.73±2.62 |
| **TSDDNet (Ours)** | **89.62±1.24** | **90.73±1.15** | **87.59±1.27** | **78.32±1.95** |

In Fig. 7, we also present the ROC curves and AUC values for all algorithms. It is seen that the ROC curve of TSDDNet outperforms all other algorithms with the highest AUC value of 0.928, indicating the best classification performance.

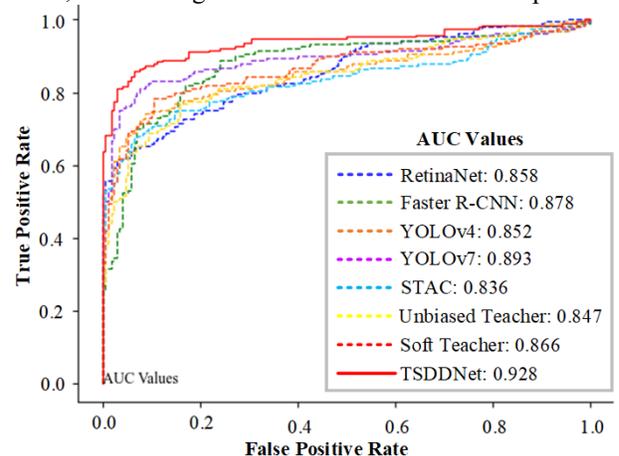

Fig. 7 ROC curves and AUC values of different algorithms on BBUI dataset

## C. Ablation Experiments

Table II shows the results of ablation experiments. It can be observed that the proposed TSDDNet improves by at least 1.70%, 2.45%, 1.58%, and 4.04% on classification accuracy, sensitivity, specificity, YI, respectively, indicating the effectiveness of distinguishing the benign and malignant nature of lesions. Moreover, the TSDDNet-B+CS improves by 4.23%, 4.52%, 4.84%, 3.65%, and 8.16% on the corresponding indices over TSDDNet-B, which suggests that the candidate selection mechanism can effectively improve the classification performance by refining ROI bounding boxes in BUS images. On the other hand, compared to TSDDNet-B, TSDDNet-B+SD achieves improvements of 1.36%, 3.03%, 2.04%, and 5.07% on accuracy, sensitivity, specificity, YI, respectively. This fully demonstrates the effectiveness of F-Net, which transfers knowledge to D-Net and C-Net via self-distillation, helping both networks to learn more discriminant features for classification. While comparing the TSDDNet-B+CS with TSDDNet-B+SD, it can be found that the former outperforms the latter, suggesting the candidate selection plays more important role for superior performance. From Fig. 8, it can also be observed that by the introducing the candidate selection mechanism, the predicted ROIs are more reasonable compared to ground truth, and the classification probabilities is also higher. This indicates that the refined ROIs are more suitable for the classification task.

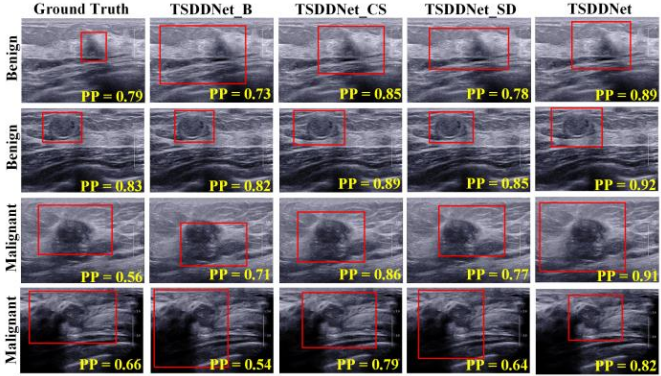

Fig. 8 Visualization results of ablation experiments. PP means Predicted Probability.

Fig. shows the classification performance of TSDDNet and TSDDNet-B+SD in the first and second stages. Compared with TSDDNet, TSDDNet-B+SD removed the candidate selection mechanism, which mean that TSDDNet-B+SD did not refine manual ROI-level labels during training. It was worth noting that the annotated ROI ratio $p$ was set to 1, which indicated that all samples in the training set were annotated with manual ROIs. In the first stage, the classification performance of TSDDNet improves by 2.74%, 2.85%, 2.40% and 5.25% on classification accuracy, sensitivity, specificity, and YI, respectively, over TSDDNet-B+SD. In the second stage, TSDDNet also performs better than TSDDNet-B+SD, which improves by 2.58%, 1.33%, 3.70% and 5.02% on accuracy, sensitivity, specificity, and YI. These observations indicate that manually ROI-level annotation is not optimal for the subsequent classification and can be refined to be more suitable for the classification tasks.

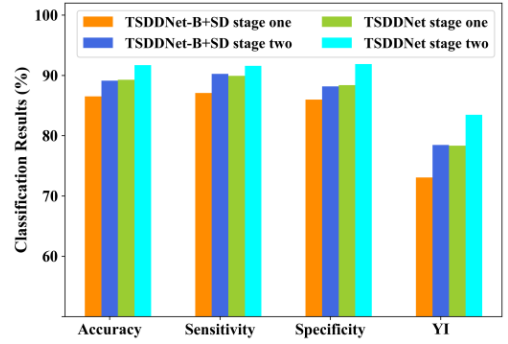

Fig. 9 Results of refined ROI-level labels.

## D. Analysis on Different $p$

To control the percentage of the samples with ROI-level labels in the training set, the parameter $p$ was introduced. In previous experiments, the $p$ for weakly supervised algorithms was set to 0.2, which indicated that 20% of the patient samples in the training set were randomly selected to retain the ROI-level labels. 80% of the ROI-level annotating effort was saved by giving only class labels.

The following experiments were conducted to compare the classification performance of our TSDDNet when $p$ value varied. In Table III, the classification results of STAC and our TSDDNet with different $p$ are presented, including $p = 0.2$, $p = 0.4$, $p = 0.6$, $p = 0.8$. Obviously, the proposed TSDDNet outperforms the WSL-based algorithm STAC and supervised learning-based algorithm RetinaNet, respectively. It demonstrates that our TSDDNet can efficiently alleviate the issue of SSS and coarse annotation and significantly improve the classification results.

TABLE II
ABLATION EXPERIMENT ON BBUI DATASET (UNIT: %)

|  | Accuracy | Sensitivity | Specificity | YI |
|---|---|---|---|---|
| RetinaNet | 84.84±1.08 | 86.01±0.82 | 82.70±1.09 | 68.71±1.74 |
| TSDDNet-B | 83.69±1.61 | 83.76±1.21 | 82.36±1.10 | 66.12±1.85 |
| TSDDNet-B+CS | 87.92±1.77 | 88.28±1.62 | 86.01±2.10 | 74.28±2.08 |
| TSDDNet-B+SD | 85.05±1.14 | 86.79±1.45 | 84.40±1.43 | 71.19±2.30 |
| **TSDDNet** | **89.62±1.24** | **90.73±1.15** | **87.59±1.27** | **78.32±1.95** |

TABLE III
CLASSIFICATION RESULTS OF DIFFERENT $p$ ON BBUI DATASET (UNIT: %)

|  | $p$ | Accuracy | Sensitivity | Specificity | YI |
|---|---|---|---|---|---|
| RetinaNet | 1.0 | 84.84±1.08 | 86.01±0.82 | 82.70±1.09 | 68.71±1.74 |
| STAC | 0.2 | 72.42±1.70 | 83.33±1.13 | 81.34±1.07 | 64.67±1.99 |
| **TSDDNet** |  | **89.62±1.24** | **90.73±1.15** | **87.59±1.27** | **78.32±1.95** |
| STAC | 0.4 | 74.85±1.42 | 83.82±1.36 | 81.10±1.02 | 64.92±2.14 |
| **TSDDNet** |  | **90.19±1.30** | **91.26±1.31** | **88.62±1.09** | **79.88±2.18** |
| STAC | 0.6 | 78.48±1.26 | 84.43±1.27 | 82.06±1.08 | 66.49±2.25 |
| **TSDDNet** |  | **91.06±1.18** | **92.15±1.32** | **89.41±1.27** | **81.66±2.16** |
| STAC | 0.8 | 82.42±1.70 | 83.33±1.13 | 81.34±1.07 | 64.67±1.99 |
| **TSDDNet** |  | **91.57±1.23** | **92.40±1.13** | **90.76±1.08** | **83.16±2.09** |

It can be observed from Table III that the performance of our TSDDNet goes steady when the parameter $p$ varies, indicating that the performance of TSDDNet is insensitive to the parameter $p$. The main reason is that our TSDDNet adopted the candidate selection mechanism, which refined ROIs according to the candidate classification result during the training. Intrinsically, this candidate selection mechanism can reduce the negative impact of coarse annotation on the model training and constrain the ROI-level labels towards higher classification performance. Therefore, it also enhances the robustness of the TSDDNet.

## VI. DISCUSSION

In this work, a WSL-based TSDDNet is proposed for BUS-based CAD, in which two different networks, i.e., lesion detection network and classification network, are integrated into a unified CAD framework to automatically detect and diagnose breast cancers from BUS images. Meanwhile, a two-stage training strategy is designed to improve detection and classification accuracy with both the partially and fully annotated training samples and refine the ROI-level labels. The experimental results on BUS image datasets indicate the effectiveness of the proposed TSDDNet.

In clinical practice, existing DL-based CAD for breast cancers still has some limitations. For example, the collection of BUS images with ROI-level labels is time-consuming and laborious [5]. Consequently, the DL model cannot be well trained with limited training samples.

Apart from the issue of SSS, existing lesion detection methods still cannot handle the issue of coarse annotation. This is due to the personal experience of different sonographers, the quality of the annotated ROI bounding boxes of BUS images is uneven, some of them may not be the best regions for the classification task.

To solve the abovementioned issues of SSS and coarse annotation, we label the BUS images without ROI-level annotation and design a candidate selection mechanism to refine coarse annotations. To be specifically, the BUS images with both image-level and ROI-level annotation are fed to train D-Net and C-Net. Then trained D-Net is used to generate pseudo-ROI-level annotation for images lacking them, which can alleviate the issue of SSS and reduce the time consumption of manual annotation. In addition, the pseudo-ROI-level annotation and ground truth can be optimized during the first stage due to the candidate selection mechanism. In detail, The BUS images are fed to the D-Net to predict k ROI candidates during k iterations and the C-Net evaluates the probabilities on each candidate, selecting the candidate with the highest probability as the new ROI-level label.

Different from the criterion for ROI annotation in [11] that obtained a certain proportion constant 0.6 through experiments, we designed a candidate selection mechanism for the network to annotate each BUS image with a suitable bounding box automatically. Because of uniqueness of each BUS image, it is unreasonable to set a fixed size for all BUS images, which will result in missing or redundant information in some images. Therefore, each BUS image should receive its own unique bounding box size to participate in network training.

Although our proposed TSDDNet has achieved remarkable results for BUS-based CAD, it still can be improved. For example, the proposed two-stage training strategy is effective, but fine-tuning the hyperparameters of the network could be a time-consuming and complex process. We will consider about integrating two stages into one stage for convenience in our future work. Furthermore, the mechanism of iteratively refining ROI through classification probabilities has improved classification performance, but its optimization efficiency is low, requiring multiple iterations to achieve better results. A more efficient method of refining ROI will be explored in the subsequent research.

## VII. CONCLUSION

In conclusion, a novel TSDDNet is proposed to automatically detect and diagnose breast cancers, which is trained using only coarsely and partially annotated ROIs in BUS images. It integrates lesion detection network and classification network into a unified CAD framework, and a two-stage training strategy is developed to solve the issue of coarse annotation together with the SSS problem. Specifically, the WSL-based ROI refinement method can effectively refine manually annotated ground truth ROIs, which is beneficial to the training of detection and classification models. Extensive experiments indicate that the proposed TSDDNet outperforms all compared algorithms, indicating its potential applications.